\documentclass[twocolumn,trackchanges]{aastex701}

\usepackage{enumitem}

\begin{document}

\title{FAST Burst-Cluster Catalog of FRB~20240114A}

\author[orcid=0009-0002-3020-9123]{Long-Xuan Zhang}
\affiliation{School of Physics, Huazhong University of Science and Technology, Wuhan 430074, China}
\email{d202580060@hust.edu.cn}

\author[orcid=0009-0002-7318-435X]{Ya Zeng}
\affiliation{School of Physics, Huazhong University of Science and Technology, Wuhan 430074, China}
\email{2739706499@qq.com}

\author[orcid=0009-0008-4576-3120]{Tong-Lun Wang}
\affiliation{School of Physics, Huazhong University of Science and Technology, Wuhan 430074, China}
\email{2464601354@qq.com}

\author[orcid=0000-0002-7949-3906]{Jun-Yi Shen}
\affiliation{School of Physics, Huazhong University of Science and Technology, Wuhan 430074, China}
\email{309238461@qq.com}

\author[orcid=0009-0004-5530-4447]{Ying-Ze Shan}
\affiliation{School of Physics, Huazhong University of Science and Technology, Wuhan 430074, China}
\email{d202480059@hust.edu.cn}

\author[orcid=0009-0004-8180-3055]{Lin Zhou}
\affiliation{School of Physics, Huazhong University of Science and Technology, Wuhan 430074, China}
\email{1822548678@qq.com}

\author{Shi-Yan Tian}
\affiliation{School of Physics, Huazhong University of Science and Technology, Wuhan 430074, China}
\email{2226247263@qq.com}


\author[orcid=0000-0002-9332-5562]{Tian-Cong Wang}
\affiliation{School of Physics and Astronomy, Beijing Normal University, Beijing 100875, China}
\affiliation{Institute for Frontiers in Astronomy and Astrophysics, Beijing Normal University, Beijing 102206, China}
\email{wangtc@mail.bnu.edu.cn}

\author[orcid=0000-0002-7420-9988]{Deng-Ke Zhou}
\affiliation{Research Center for Astronomical Computing, Zhejiang Laboratory, Hangzhou 311121, China}
\email{zdk@zhejianglab.com}

\author[orcid=0000-0002-6423-6106]{De-Jiang Zhou}
\affiliation{National Astronomical Observatories, Chinese Academy of Sciences, Beijing 100101, China}
\email{djzhou@nao.cas.cn}

\author[orcid=0000-0002-9274-3092]{Jin-Lin Han}
\affiliation{National Astronomical Observatories, Chinese Academy of Sciences, Beijing 100101, China}
\affiliation{University of Chinese Academy of Sciences, Beijing 100049, China}
\affiliation{State Key Laboratory of Radio Astronomy and Technology, Beijing 100101, China}
\email{hjl@nao.cas.cn}

\author[orcid=0000-0001-8065-4191]{Jia-Rui Niu}
\affiliation{National Astronomical Observatories, Chinese Academy of Sciences, Beijing 100101, China}
\email{niujiarui@nao.cas.cn}

\author[orcid=0009-0007-3326-7827]{Rui-Nan Li}
\affiliation{School of Astronomy and Space Science, Nanjing University, Nanjing 210023, China}
\email{liruinan927@smail.nju.edu.cn}

\author[orcid=0009-0000-7501-2215]{Jin-Huang Cao}
\affiliation{National Astronomical Observatories, Chinese Academy of Sciences, Beijing 100101, China}
\affiliation{University of Chinese Academy of Sciences, Beijing 100049, China}
\email{caojinhuang22@mails.ucas.ac.cn}

\author[orcid=0000-0002-4300-121X]{Rui Luo}
\affiliation{Department of Astronomy, School of Physics and Materials Science, Guangzhou University, Guangzhou 510006, China}
\email{rui.luo@gzhu.edu.cn}

\author[orcid=0000-0002-6465-0091]{Jin-Chen Jiang}
\affiliation{Max-Planck-Institut für Radioastronomie, Auf dem Hügel 69, Bonn, D-53121, Germany}
\affiliation{National Astronomical Observatories, Chinese Academy of Sciences, Beijing 100101, China}
\email{jjiang@mpifr-bonn.mpg.de}

\author[orcid=0000-0002-6165-0977]{Xiang-Han Cui}
\affiliation{National Astronomical Observatories, Chinese Academy of Sciences, Beijing 100101, China}
\email{cuixianghan@nao.cas.cn}

\author[orcid=0000-0001-9036-8543]{Wei-Yang Wang}
\affiliation{University of Chinese Academy of Sciences, Beijing 100049, China}
\email{wywang_astroph@pku.edu.cn}

\author[orcid=0009-0008-6247-0645]{Wen-Long Zhang}
\affiliation{Purple Mountain Observatory, Chinese Academy of Sciences, Nanjing 210023, China}
\affiliation{School of Astronomy and Space Science, University of Science and Technology of China, Hefei 230026, China}
\email{wlzhang@pmo.ac.cn}

\author[orcid=0009-0008-5254-7792]{Si-Lu Xu}
\affiliation{National Astronomical Observatories, Chinese Academy of Sciences, Beijing 100101, China}
\affiliation{University of Chinese Academy of Sciences, Beijing 100049, China}
\email{xusilu@buaa.edu.cn}

\author[orcid=0000-0002-3386-7159]{Pei Wang}
\affiliation{National Astronomical Observatories, Chinese Academy of Sciences, Beijing 100101, China}
\affiliation{Institute for Frontiers in Astronomy and Astrophysics, Beijing Normal University, Beijing 102206, China}
\affiliation{State Key Laboratory of Radio Astronomy and Technology, Beijing 100101, China}
\email{wangpei@nao.cas.cn} 

\correspondingauthor{Pei Wang}
\email{wangpei@nao.cas.cn}

\author[orcid=0000-0001-5105-4058]{Wei-Wei Zhu}
\affiliation{National Astronomical Observatories, Chinese Academy of Sciences, Beijing 100101, China}
\affiliation{Institute for Frontiers in Astronomy and Astrophysics, Beijing Normal University, Beijing 102206, China}
\affiliation{State Key Laboratory of Radio Astronomy and Technology, Beijing 100101, China}
\email{zhuww@nao.cas.cn}

\correspondingauthor{Wei-Wei Zhu}
\email{zhuww@nao.cas.cn}

\author[orcid=0000-0002-9725-2524]{Bing Zhang}
\affiliation{The Hong Kong Institute for Astronomy and Astrophysics, the University of Hong Kong, Pokfulam, Hong Kong, China}
\affiliation{Department of Physics, the University of Hong Kong, Pokfulam, Hong Kong, China}
\affiliation{The Nevada Center for Astrophysics, University of Nevada, Las Vegas, Las Vegas, NV 89154, USA}
\affiliation{Department of Physics and Astronomy, University of Nevada, Las Vegas, Las Vegas, NV 89154, USA}
\email{bzhang1@hku.hk}  

\correspondingauthor{Bing Zhang}
\email{bzhang1@hku.hk}  

\author[orcid=0000-0003-3010-7661]{Di Li}
\affiliation{Department of Astronomy, Tsinghua University, Beijing 100084, China}
\affiliation{National Astronomical Observatories, Chinese Academy of Sciences, Beijing 100101, China}
\affiliation{State Key Laboratory of Radio Astronomy and Technology, Beijing 100101, China}
\email{dili@tsinghua.edu.cn}

\correspondingauthor{Di Li}
\email{dili@tsinghua.edu.cn}

\author[orcid=0000-0003-4137-4247]{Yun-Peng Men}
\affiliation{School of Physics, Huazhong University of Science and Technology, Wuhan 430074, China}
\affiliation{Hubei Fundamental Research Center for Physics, Wuhan 430074, China}
\affiliation{Hubei Key Laboratory of Gravitation and Quantum Physics, Huazhong University of Science and Technology, Wuhan 430074, China}
\affiliation{Max-Planck-Institut für Radioastronomie, Auf dem Hügel 69, Bonn, D-53121, Germany}
\email{ypmen@hust.edu.cn}

\correspondingauthor{Yun-Peng Men}
\email{ypmen@hust.edu.cn}

\author[orcid=0000-0002-5400-3261]{Yuan-Chuan Zou}
\affiliation{School of Physics, Huazhong University of Science and Technology, Wuhan 430074, China}
\affiliation{Hubei Fundamental Research Center for Physics, Wuhan 430074, China}
\affiliation{Hubei Key Laboratory of Gravitation and Quantum Physics, Huazhong University of Science and Technology, Wuhan 430074, China}
\affiliation{Purple Mountain Observatory, Chinese Academy of Sciences, Nanjing 210023, China}
\email{zouyc@hust.edu.cn}

\correspondingauthor{Yuan-Chuan Zou}
\email{zouyc@hust.edu.cn}

\author[orcid=0000-0001-5950-7170]{Wan-Peng Sun}
\affiliation{Liaoning Key Laboratory of Cosmology and Astrophysics, College of Sciences, Northeastern University, Shenyang 110819, China}
\email{sunwp@mails.neu.edu.cn}

\author[orcid=0000-0002-6029-1933]{Xin Zhang}
\affiliation{Liaoning Key Laboratory of Cosmology and Astrophysics, College of Sciences, Northeastern University, Shenyang 110819, China}
\affiliation{MOE Key Laboratory of Data Analytics and Optimization for Smart Industry, Northeastern University, Shenyang 110819, China}
\affiliation{National Frontiers Science Center for Industrial Intelligence and Systems Optimization, Northeastern University, Shenyang 110819, China}
\email{zhangxin@neu.edu.cn}

\author[orcid=0000-0001-6651-7799]{Chen-Hui Niu}
\affiliation{Institute of Astrophysics, Central China Normal University, Wuhan 430079, China}
\email{niuchenhui@ccnu.edu.cn}

\author[orcid=0000-0001-5931-2381]{Ye Li}
\affiliation{Purple Mountain Observatory, Chinese Academy of Sciences, Nanjing 210023, China}
\email{yeli@pmo.ac.cn}

\author[orcid=0000-0003-4721-4869]{Yuan-Hong Qu}
\affiliation{Department of Physics, University of Helsinki, P.O. Box 64, FI-00014 University of Helsinki, Finland}
\email{yuanhong.qu@unlv.edu}

\author[orcid=0000-0002-2552-7277]{Xiao-Hui Liu}
\affiliation{State Key Laboratory of Radio Astronomy and Technology, Beijing 100101, China}
\affiliation{University of Chinese Academy of Sciences, Beijing 100049, China}
\email{liuxh@bao.ac.cn}

\author[orcid=0000-0001-6374-8313]{Yuan-Pei Yang}
\affiliation{South-Western Institute for Astronomy Research, Key Laboratory of Survey Science of Yunnan Province, Yunnan University, Kunming 650500, China}
\email{ypyang@ynu.edu.cn}

\author[orcid=0000-0002-1435-0883]{Ke-Jia Lee}
\affiliation{Department of Astronomy, School of physics, Peking University, Beijing 100871, China}
\affiliation{National Astronomical Observatories, Chinese Academy of Sciences, Beijing 100101, China}
\affiliation{Yunnan Astronomical Observatories, Chinese Academy of Sciences, Kunming 650216, China}
\affiliation{Beijing Laser Acceleration Innovation Center, Huairou, Beijing 101400, China}
\email{kjlee@pku.edu.cn}

\author[orcid=0000-0002-9642-9682]{Jia-Wei Luo}
\affiliation{College of Physics and Hebei Key Laboratory of Photophysics Research and Application, Hebei Normal University, Shijiazhuang 050024, China}
\affiliation{Shijiazhuang Key Laboratory of Astronomy and Space Science, Hebei Normal University, Shijiazhuang 050024, China}
\email{ljw@hebtu.edu.cn}

\author[orcid=0000-0002-0475-7479]{Yi Feng}
\affiliation{Research Center for Astronomical Computing, Zhejiang Laboratory, Hangzhou 311121, China}
\affiliation{Institute for Astronomy, School of Physics, Zhejiang University, Hangzhou 310027, China}
\email{yifeng@zhejianglab.org}

\author[orcid=0000-0001-7931-0607]{Dong-Zi Li}
\affiliation{Department of Astronomy, Tsinghua University, Beijing 100084, China}
\email{dzli@tsinghua.edu.cn}

\author[orcid=0000-0001-7199-2906]{Yong-Feng Huang}
\affiliation{School of Astronomy and Space Science, Nanjing University, Nanjing 210023, China}
\affiliation{Key Laboratory of Modern Astronomy and Astrophysics (Nanjing University), Ministry of Education, Nanjing 210093, China}
\email{hyf@nju.edu.cn}

\author[orcid=0000-0002-3844-9677]{Wen-Fei Yu}
\affiliation{Shanghai Astronomical Observatory, Chinese Academy of Sciences, Shanghai 200030, China}
\email{wenfei@shao.ac.cn}

\author[orcid=0000-0001-6021-5933]{Qin Wu}
\affiliation{School of Astronomy and Space Science, Nanjing University, Nanjing 210023, China}
\email{wqin@smail.nju.edu.cn}

\author[orcid=0000-0002-3100-6558]{He Gao}
\affiliation{School of Physics and Astronomy, Beijing Normal University, Beijing 100875, China}
\affiliation{Institute for Frontiers in Astronomy and Astrophysics, Beijing Normal University, Beijing 102206, China}
\email{gaohe@bnu.edu.cn}

\author[orcid=0000-0003-4157-7714]{Fa-Yin Wang}
\affiliation{School of Astronomy and Space Science, Nanjing University, Nanjing 210023, China}
\affiliation{Key Laboratory of Modern Astronomy and Astrophysics (Nanjing University), Ministry of Education, Nanjing 210093, China}
\email{fayinwang@nju.edu.cn}


\author[orcid=0000-0002-1056-5895]{Wei-Cong Jing}
\affiliation{National Astronomical Observatories, Chinese Academy of Sciences, Beijing 100101, China}
\email{jingweicong@nao.cas.cn}

\author[orcid=0000-0002-9390-9672]{Chao-Wei Tsai}
\affiliation{National Astronomical Observatories, Chinese Academy of Sciences, Beijing 100101, China}
\affiliation{Institute for Frontiers in Astronomy and Astrophysics, Beijing Normal University,  Beijing 102206, China}
\affiliation{University of Chinese Academy of Sciences, Beijing 100049, China}
\affiliation{State Key Laboratory of Radio Astronomy and Technology, Beijing 100101, China}
\email{cwtsai@nao.cas.cn}

\author[orcid=0000-0002-0633-5325]{Lin Lin}
\affiliation{School of Physics and Astronomy, Beijing Normal University, Beijing 100875, China}
\email{llin@bnu.edu.cn}

\author[orcid=0000-0001-5649-2591]{Jin-Tao Xie}
\affiliation{School of Computer Science and Engineering, Sichuan University of Science and Engineering, Yibin 644000, China}
\email{xiejintao@suse.edu.cn}

\author[orcid=0000-0001-9956-6298]{Jian-Hua Fang}
\affiliation{Research Center for Astronomical Computing, Zhejiang Laboratory, Hangzhou 311121, China}
\email{fangjh@zhejianglab.com}

\author[orcid=0000-0002-9441-2190]{Chen-Chen Miao}
\affiliation{College of Physics and Electronic Engineering, Qilu Normal University, Jinan 250200, China}
\email{miaocc@nao.cas.cn}

\author[orcid=0009-0009-8320-1484]{Yu-Hao Zhu}
\affiliation{National Astronomical Observatories, Chinese Academy of Sciences, Beijing 100101, China}
\affiliation{University of Chinese Academy of Sciences, Beijing 100049, China}
\email{zhuyh@bao.ac.cn}

\author[orcid=0009-0000-4795-8767]{Yun-Chuan Chen}
\affiliation{Research Center for Astronomical Computing, Zhejiang Laboratory, Hangzhou 311121, China}
\email{chenych@zhejianglab.org}

\author[orcid=0000-0002-8744-3546]{Yong-Kun Zhang}
\affiliation{National Astronomical Observatories, Chinese Academy of Sciences, Beijing 100101, China}
\email{ykzhang@nao.cas.cn}

\author[orcid=0000-0002-8870-981X]{Shuo Cao}
\affiliation{University of Chinese Academy of Sciences, Beijing 100049, China}
\email{caoshuo@bao.ac.cn}

\author[orcid=0000-0002-1381-7859]{Zi-Wei Wu}
\affiliation{National Astronomical Observatories, Chinese Academy of Sciences, Beijing 100101, China}
\email{wuzw@bao.ac.cn}

\author[orcid=0000-0002-4327-711X]{Chun-Feng Zhang}
\affiliation{National Astronomical Observatories, Chinese Academy of Sciences, Beijing 100101, China}
\email{cf.zhang@pku.edu.cn}

\author[orcid=0000-0002-9434-4773]{Bo-Jun Wang}
\affiliation{National Astronomical Observatories, Chinese Academy of Sciences, Beijing 100101, China}
\email{wangbj@bao.ac.cn}

\author[orcid=0009-0000-6108-2730]{Hua-Xi Chen}
\affiliation{Research Center for Astronomical Computing, Zhejiang Laboratory, Hangzhou 311121, China}
\email{chenhxtony@gmail.com}

\author[orcid=0000-0001-5738-9625]{Xiang-Lei Chen}
\affiliation{National Astronomical Observatories, Chinese Academy of Sciences, Beijing 100101, China}
\email{xiangleichen@nao.cas.cn}

\author[orcid=0009-0000-8142-8612]{Yu-Xiang Huang}
\affiliation{Yunnan Astronomical Observatories, Chinese Academy of Sciences, Kunming 650216, China}
\email{huangyuxiang@ynao.ac.cn}

\author[orcid=0000-0003-1720-9727]{Jian Li}
\affiliation{Department of Astronomy, University of Science and Technology of China, Hefei 230026, China}
\affiliation{School of Astronomy and Space Science, University of Science and Technology of China, Hefei 230026, China}
\email{jianli@ustc.edu.cn}

\author[orcid=0009-0005-5413-7664]{Qing-Yue Qu}
\affiliation{National Astronomical Observatories, Chinese Academy of Sciences, Beijing 100101, China}
\affiliation{University of Chinese Academy of Sciences, Beijing 100049, China}
\email{quqy@bao.ac.cn}

\author[orcid=0000-0002-7372-4160]{Yi-Dan Wang}
\affiliation{National Astronomical Observatories, Chinese Academy of Sciences, Beijing 100101, China}
\affiliation{University of Chinese Academy of Sciences, Beijing 100049, China}
\email{wangyidan19@mails.ucas.ac.cn}

\author[orcid=0000-0001-7746-9462]{Su-Ming Weng}
\affiliation{National Key Laboratory of Dark Matter Physics, School of Physics and Astronomy, Shanghai Jiao Tong University, Shanghai 200240, China}
\affiliation{Laboratory for Laser Plasmas and Collaborative Innovation Centre of IFSA, Shanghai Jiao Tong University, Shanghai 200240, China} 
\email{wengsuming@sjtu.edu.cn}

\author[orcid=0000-0002-6299-1263]{Xue-Feng Wu}
\affiliation{Purple Mountain Observatory, Chinese Academy of Sciences, Nanjing 210023, China}
\email{xfwu@pmo.ac.cn}

\author[orcid=0000-0002-5799-9869]{Shi-Han Yew}
\affiliation{National Key Laboratory of Dark Matter Physics, School of Physics and Astronomy, Shanghai Jiao Tong University, Shanghai 200240, China}
\affiliation{Laboratory for Laser Plasmas and Collaborative Innovation Centre of IFSA, Shanghai Jiao Tong University, Shanghai 200240, China}
\email{yushihan@sjtu.edu.cn}

\author[orcid=0000-0003-4546-2623]{Ai-Yuan Yang}
\affiliation{National Astronomical Observatories, Chinese Academy of Sciences, Beijing 100101, China}
\affiliation{State Key Laboratory of Radio Astronomy and Technology, Beijing 100101, China}
\email{yangay@bao.ac.cn}

\author[orcid=0000-0001-8539-4237]{Lei Zhang}
\affiliation{National Astronomical Observatories, Chinese Academy of Sciences, Beijing 100101, China}
\affiliation{Centre for Astrophysics and Supercomputing, Swinburne University of Technology, Hawthorn, VIC 3122, Australia}
\email{leizhang996@nao.cas.cn}

\author[orcid=0000-0002-1243-0476]{Ru-Shuang Zhao}
\affiliation{Guizhou Provincial Key Laboratory of Radio Astronomy and Data Processing, Guizhou Normal University, Guiyang 550001, China}
\email{201907007@gznu.edu.cn}

\collaboration{all}{The FAST FRB KSP Collaborations.}
\begin{abstract}
We present the complete FAST burst-cluster catalog of FRB~20240114A, covering its 495-day active episode from January~28, 2024 to June~5, 2025.
The catalog contains 13261 burst clusters, 19037 bursts, and 29026 sub-bursts detected in 99 observing sessions.
The data reveal distinct organization across temporal scales.
Among drifting burst clusters, downward-drifting clusters outnumber upward-drifting ones by approximately \(9.4\!:\!1\).
Between separate burst clusters, previously reported frequency changes show no preferred direction, while our Weibull analysis finds their arrival times broadly consistent with Poisson statistics, suggesting that burst clusters may represent approximately independent emission episodes within observing sessions.
The Parkes/Murriyang observations revealed a long-term modulation of the characteristic emission frequency, and the extended FAST measurements show a consistent evolution.
Burst energetics increase with cluster multiplicity and activity level, although significant scatter remains.
The manually refined dispersion measure increases at \(1.29\pm0.11~\mathrm{pc\,cm^{-3}\,yr^{-1}}\), enlarging the still limited but growing sample of repeaters with increasing DM.
Overall, these results summarize the spectral--temporal evolution of FRB~20240114A and present a burst-cluster catalog linked to the existing BurstID measurements, providing a basis for future studies of repeating FRBs.
\end{abstract}

\keywords{
    \href{https://astrothesaurus.org/uat/2008}{Radio transient sources (2008)},
    \href{https://astrothesaurus.org/uat/1339}{Radio bursts (1339)},
    \href{https://astrothesaurus.org/uat/1338}{Radio astronomy (1338)},
    \href{https://astrothesaurus.org/uat/740}{High time resolution astrophysics (740)},
    \href{https://astrothesaurus.org/uat/739}{High energy astrophysics (739)}
}


\section{Introduction} 
\setcounter{footnote}{0}

Fast radio bursts (FRBs) are millisecond-duration radio transients characterized by large dispersion measures (DMs) and extremely high brightness temperatures, requiring coherent emission mechanisms \citep{2023RvMP...95c5005Z}. 
Comparisons between observed FRB DMs and predictions from Galactic free-electron density models, including NE2001, YMW16, and NE2025, indicate that the vast majority of known FRBs are extragalactic \citep{2002astro.ph..7156C, 2017ApJ...835...29Y, 2026ApJ..1002....3O}. 
The association of FRB~20200428 with the Galactic magnetar SGR~1935+2154 demonstrated that magnetars can also produce FRB-like radio bursts, although at radio energies much lower than typical cosmological FRBs \citep{2020Natur.587...59B, 2020Natur.587...54C}. 
Since the discovery of the first FRB source in archival Parkes data \citep{2007Sci...318..777L} and the identification of the first repeating source FRB~20121102A \citep{2014ApJ...790..101S,2016Natur.531..202S}, repeating FRBs have enabled systematic investigations of burst properties, population differences, and long-term activity evolution \citep{Luo:2022smj,Zhu-Ge:2022nkz,Sun:2024huw,Sun:2025cdq,Sun:2026aeh,2025arXiv250906208K,2020MNRAS.495.3551R,2021MNRAS.500..448C,2025A&A...693A..40B}.
More than 4500 FRB sources have been reported, including over 120 confirmed repeaters and more than 150 sources with identified host galaxies. 
These samples are continuously updated through Blinkverse\footnote{\url{https://blinkverse.zero2x.org/}} and the Transient Name Server (TNS)\footnote{\url{https://www.wis-tns.org/search}}, while CHIME \citep{2018ApJ...863...48C} provides the largest homogeneous FRB samples through its Second Catalog \citep{2026arXiv260109399F}. 
The sensitivity of FAST \citep{2011IJMPD..20..989N,2019SCPMA..6259502J,2020Innov...100053Q} enables detailed long-term monitoring of repeating sources.

Among the growing population of known FRBs, FRB~20240114A has emerged as an exceptionally active repeater, with its discovery by CHIME revealing a high burst rate and large dispersion measure \citep{2026ApJ...997..334S,2023ApJ...947...83C}. 
Subsequent multi-facility observations, including MeerKAT, EVN/PRECISE, Gran Telescopio Canarias, and uGMRT, have established its precise localization, host-galaxy association, and broadband burst activity \citep{2024MNRAS.533.3174T,2025ApJ...992L..35B,2025ApJ...989...15P}. 
Radio-continuum and high-resolution observations with MeerKAT, uGMRT, e-MERLIN, and VLBA revealed a persistent radio source associated with FRB~20240114A and its long-term variability, suggesting a possible connection between persistent emission and burst activity \citep{2025A&A...695L..12B,2025arXiv250114247Z,2026ApJ..1004..101L}. 
Extensive monitoring with GBT, Effelsberg, TMRT, the Allen Telescope Array, and Parkes/Murriyang further characterized the burst activity, polarization properties, frequency-dependent emission, and possible propagation effects of this source \citep{2025ApJS..278...49X,2026A&A...709A..47L,2026arXiv260428012M,2025A&A...695L..10E,2025ApJ...992..185W,2026arXiv260631897J,2026arXiv260216409U,2026arXiv260512098L,2026arXiv260702939N}.
The extreme activity of FRB~20240114A was further demonstrated by the detection of an exceptionally energetic burst during a long-term monitoring campaign \citep{2026arXiv260518513O}, establishing this source as an important laboratory for studying FRB emission, propagation, and environments.

In this context, the FAST FRB Key Science Project has provided the most burst-rich monitoring dataset of FRB~20240114A to date, including a peak burst rate exceeding \(700~\mathrm{hr}^{-1}\) during the March~12,~2024 observation and enabling studies of the source activity, energy budget, periodicity, polarization, rotation-measure evolution, and burst-cluster morphology \citep{2025arXiv250714707Z,2026SCPMA..6949512Z,2026ApJS..284...77W,2026ApJ...998..276Z}. 
Recent statistical studies of multicomponent burst structures in repeating FRBs have revealed characteristic distributions of component numbers, suggesting that burst morphology may provide clues to the underlying emission processes \citep{2026ApJ...996....4Z}.
Independent analyses of the March~12,~2024 data found no significant periodic signal in burst arrival times, while combined KM40M and FAST observations suggested possible high-energy deviations in the cumulative burst-energy distribution \citep{2026OJAp....964830K,2025RAA....25h5009H}.

we present the FAST monitoring dataset covering the main observed active phase of FRB~20240114A and construct a burst-cluster catalog of FRB~20240114A and investigate the organization of burst emission across temporal scales and activity states, including burst-cluster evolution, long-term frequency changes, and energetic properties.
The paper is organized as follows: Section~\ref{sec:obs_datareduction} describes the observations and catalog construction; Sections~\ref{sec:long_term_evolution} and \ref{sec:energetics} characterize the burst emission properties, such as temporal, spectral, morphological, and energetic behaviors; and Section~\ref{sec:Discussion} discusses the implications and summarizes our conclusions.

\section{Observations and Data Reduction}
\label{sec:obs_datareduction}

Following its discovery by CHIME on January~14, 2024, FRB~20240114A was intensively monitored with FAST using the central beam of the L-band 19-beam receiver \citep{2018IMMag..19..112L,2020RAA....20...64J}.
All times and dates in this paper are expressed in Coordinated Universal Time (UTC).
The FAST observations adopted the initial CHIME localization and subsequently updated the pointing position following the refined MeerKAT localization \citep{2024ATel16420....1S,10.1093/mnras/stae2013}.
The monitoring campaign started on January~28, 2024 and has included more than 140 independent observing sessions to date.
During the active period from January~28, 2024 to June~5, 2025 (495 days), FAST obtained 99 valid observing sessions (98 observing days; 58.83 hr exposure), all of which yielded burst detections, resulting in 18230 catalogued \texttt{BurstID} entries.
The burst-searching procedure follows \citet{2025arXiv250714707Z}, which presented a subset of the FAST dataset and the corresponding burst-energy measurements, while the complete energy measurements will be reported in a forthcoming collaborative study by J.-S.~Zhang et al. 
No further bursts have been detected since June~5, 2025, and the subsequent quiescent phase has lasted for more than one year, indicating that our dataset covers the main observed active phase of FRB~20240114A.

Following \citet{2026ApJ...998..276Z}, we grouped temporally associated \texttt{BurstID} detections into burst clusters using a 400~ms threshold and extracted the corresponding dynamic-spectrum windows with the \texttt{replot\_fil} procedure within the \textsc{TransientX} framework \citep{2024A&A...683A.183M}.
For each cluster, the mean DM of its constituent \texttt{BurstID} entries, measured with the DM-power package \citep{2025arXiv250714707Z}, was used for initial extraction and dedispersion. 
The DM values were then manually refined in steps of 0.1~pc~cm$^{-3}$ for clusters with $N_{\rm sub}\geq2$ and a small subset of single-sub-burst clusters.
The optimized values were selected to maximize frequency alignment while preserving coherent burst and sub-burst structures; these refined values are referred to as Manual DM.
We further revised the segmentation according to the \texttt{Sub-Burst}, \texttt{Burst}, and \texttt{Burst Cluster} definitions of \citet{2026ApJ...998..276Z}. 
For \texttt{BurstID} entries containing multiple bursts separated by Intermittent Time Intervals, the bursts were treated separately, resulting in 13261 burst clusters, 19037 bursts, and 29026 sub-bursts; thus, the revised burst count exceeds the number of catalogued \texttt{BurstID} entries.
The released catalog retains the original \texttt{BurstID} identifiers and provides revised burst and sub-burst counts, enabling direct association with burst-energy and polarimetric measurements from the same FAST dataset \citep{2025arXiv250714707Z,2026ApJS..284...77W}.


\section{Long-Term Spectral--Temporal Evolution in the Burst-Cluster Framework}
\label{sec:long_term_evolution}

\begin{figure*}[t]
    \centering
    \includegraphics[width=\textwidth]{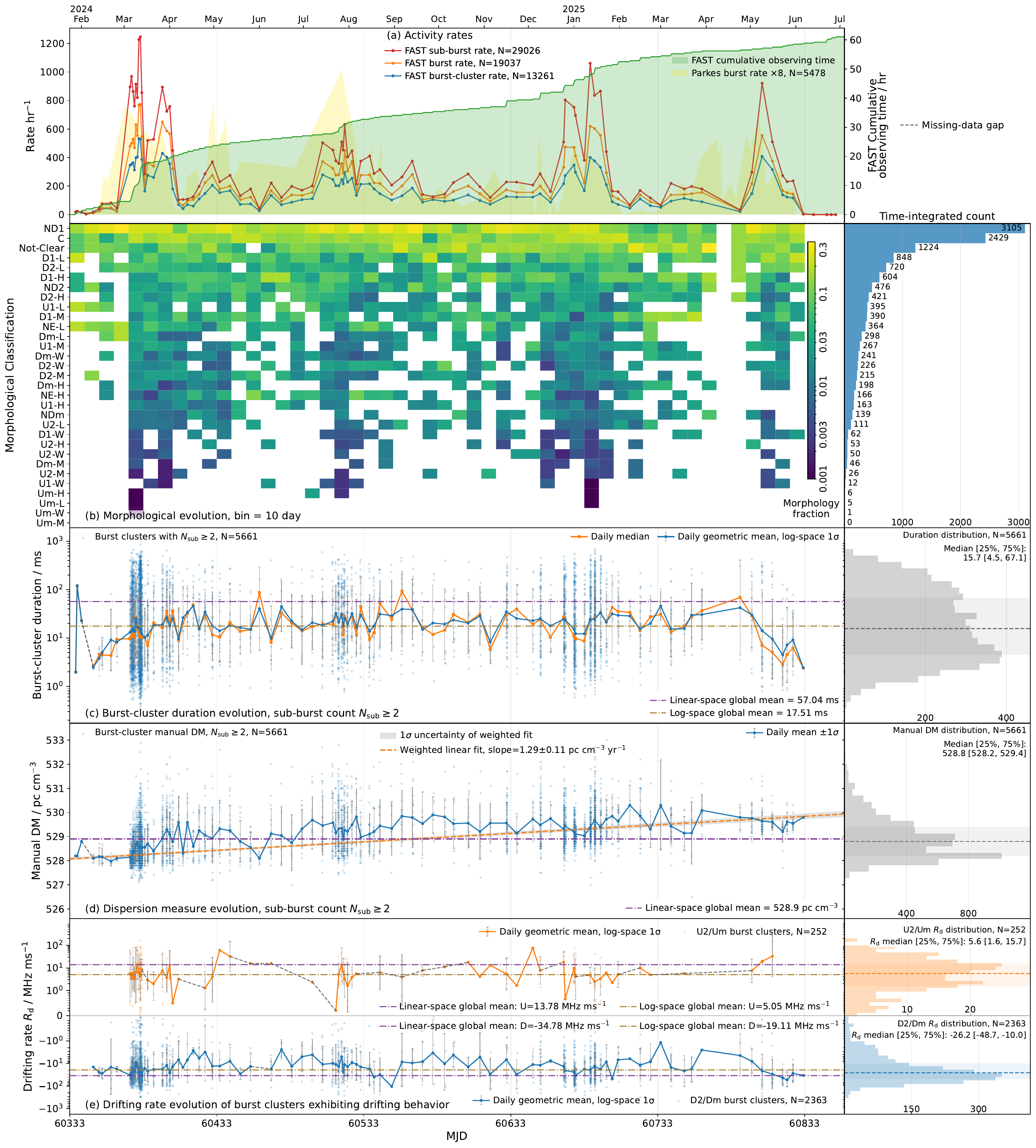}
    \caption{
    Long-term evolution of the activity, burst-cluster morphology, and other burst-cluster properties of FRB~20240114A.
    Panel (a) shows the daily FAST sub-burst, burst, and burst-cluster rates and the cumulative FAST observing time.
    For reference, the Parkes/Murriyang burst rate derived from its full-band UWL data is also shown and is scaled by a factor of eight for display \citep{2026arXiv260216409U}.
    Panel (b) shows the morphology fractions in 10-day bins, with the corresponding time-integrated class counts on the right.
    Panels (c), (d), and (e) show the duration, Manual DM, and drifting rate, respectively, for burst clusters with $N_{\rm sub}\geq2$, with panels (c) and (d) using the full sample and panel (e) only clusters exhibiting drifting behavior; the right-hand histograms show the corresponding distributions over the full observing span, with the median and 25th-75th percentile ranges indicated.
    Solid curves trace the daily statistics, grey dashed segments bridge missing-data gaps, and, in the relevant panels, purple and brown dash-dotted lines mark the linear- and log-space global means, respectively.
    }
    \label{fig:overview}
\end{figure*}

We investigate the long-term spectral--temporal evolution of FRB~20240114A using burst clusters as the basic analysis units. Figure~\ref{fig:overview} summarizes the long-term evolution of the burst activity (panel a), morphological composition (panel b), burst-cluster duration (panel c), Manual DM (panel d), and drifting rate (panel e). Since the Manual DM is fundamental for obtaining reliable burst alignment and spectral--temporal measurements, we first discuss its long-term evolution in Section~\ref{sec:dm_evolution}.

\subsection{Dispersion measure evolution and intra-day DM stability}
\label{sec:dm_evolution}

Figure~\ref{fig:overview}(d) shows the long-term evolution of the Manual DM for burst clusters with $N_{\rm sub}\geq2$.
A weighted linear fit gives $\dot{\mathrm{DM}}=1.29\pm0.11~\mathrm{pc\,cm^{-3}\,yr^{-1}}$.
This result is statistically consistent with the value of $\dot{\mathrm{DM}}=1.10\pm0.07~\mathrm{pc\,cm^{-3}\,yr^{-1}}$ reported by \citet{2026arXiv260518513O}.

\begin{figure}[t]
    \centering
    \includegraphics[width=\columnwidth]{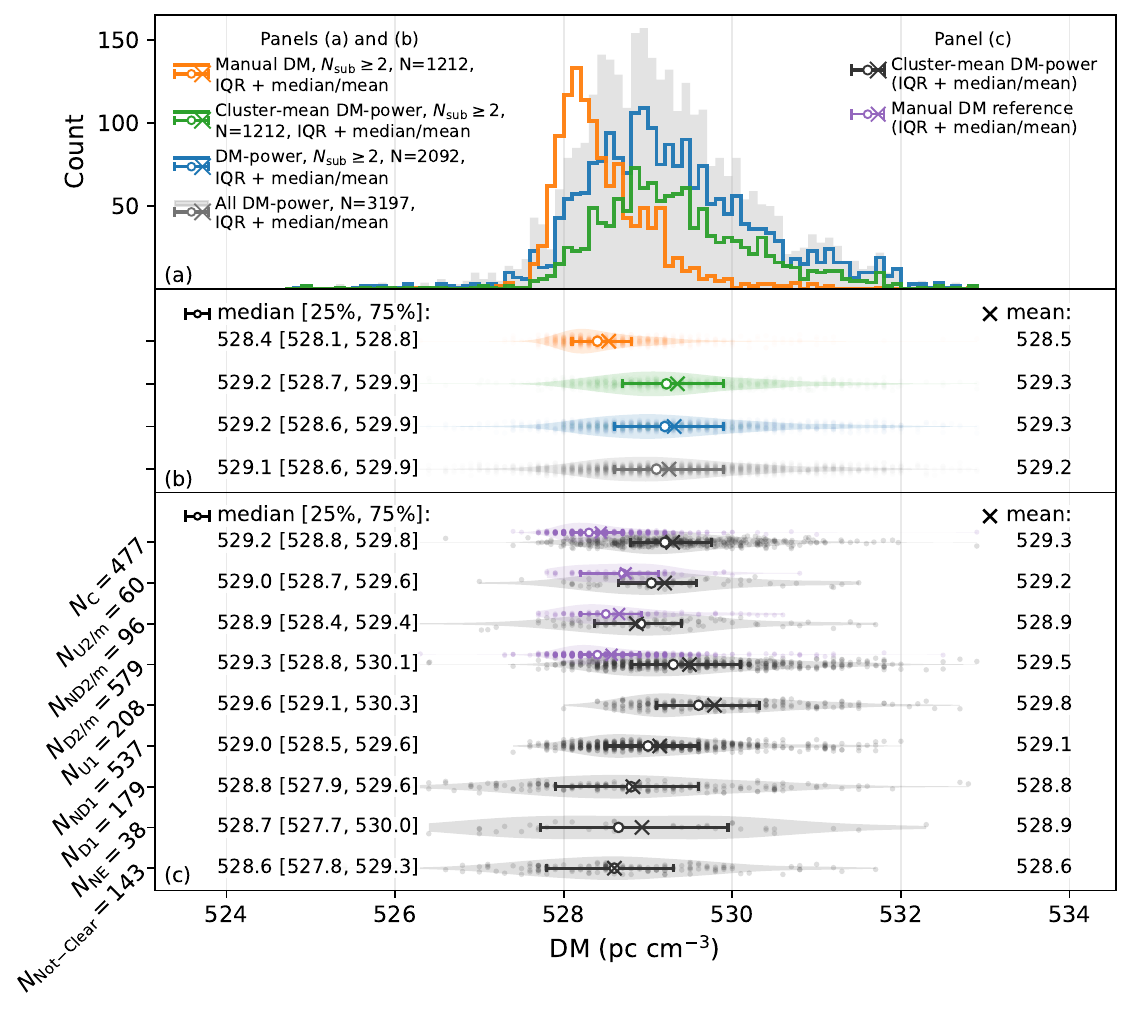}
    \caption{
    DM distributions of FRB~20240114A on 2024 March 12.
    Panels (a) and (b) compare Manual DM with cluster-mean DM-power, BurstID-level DM-power for burst clusters with $N_{\rm sub}\geq2$, and the full BurstID-level DM-power sample.
    Panel (c) compares cluster-mean DM-power and Manual DM across the merged morphological classifications.
    Horizontal bars, circles, and crosses denote the interquartile range, median, and mean, respectively.
    }
    \label{fig:dm_distribution_20240312}
\end{figure}

We further examine the intra-day DM stability using the 4.38-hr FAST observation on March~12, 2024 (Figure~\ref{fig:dm_distribution_20240312}).
For the same observing epoch, \citet{2026arXiv260703877W} compared six DM estimators and obtained a mean DM of approximately $529.2~\mathrm{pc\,cm^{-3}}$, with individual measurements exhibiting apparent burst-to-burst variations over approximately $528$--$534~\mathrm{pc\,cm^{-3}}$.
For our 1203 burst clusters with $N_{\rm sub}\geq2$, the Manual DM distribution has a mean of $528.5~\mathrm{pc\,cm^{-3}}$, a median of $528.4~\mathrm{pc\,cm^{-3}}$, and an interquartile range of $528.1$--$528.8~\mathrm{pc\,cm^{-3}}$.
The Manual DM distribution is substantially narrower than the broad burst-to-burst DM range reported by \citet{2026arXiv260703877W}, as well as the cluster-mean and BurstID-level DM-power distributions shown in Figure~\ref{fig:dm_distribution_20240312}.
In a collaborative study, D.-L.~Ma et al.\ (in preparation) applied a newly developed DM-estimation algorithm to the same observation and obtained a DM distribution closely matching our Manual DM distribution.
For another repeating source, FRB~20220912A, \citet{2026ApJ..1004..179F} used broadband microshots lasting tens of microseconds and bright narrow bursts with simple temporal profiles to argue that broad single-day DM distributions are primarily caused by morphology-induced pseudo-DM rather than rapid variations in the propagation DM.

\subsection{Long-term evolution of morphological composition and burst rate}
\label{sec:morph_rate_evolution}
Figure~\ref{fig:overview}(a) shows strong variations in the activity of FRB~20240114A throughout the monitoring campaign.
The sub-burst, burst, and burst-cluster rates evolve in parallel, with the highest activity occurring in March 2024 and additional enhancements around July 2024, January 2025, and May 2025.
The source remains detectable during quiescent intervals, indicating recurrent transitions between active and low-activity states.
The full FAST sample contains 29026 sub-bursts, 19037 bursts, and 13261 burst clusters.
Figure~\ref{fig:overview}(b) shows the temporal distribution of morphological classes in 10-day bins, revealing a possible link between activity state and burst morphology; a detailed analysis will be presented in a forthcoming companion study by L.-X.~Zhang et al.\ (in preparation).
Following the morphological classification framework described in previous studies \citep{2026ApJ...998..276Z,2022RAA....22l4001Z}, downward- and upward-drifting burst clusters are labeled in the form XY-Z, where X denotes the drifting direction (D or U), Y denotes the number of sub-bursts (1, 2, or m), and Z denotes the location of the emission within the FAST band, namely high-frequency (H), middle-frequency (M), low-frequency (L), or wide-band (W).
Other categories include non-drifting clusters (ND1, ND2, and NDm), no-evidence-for-drifting clusters (NE-H and NE-L), Not-Clear, and complex (C) clusters; representative C-class examples are provided in the public data products.

\subsection{Co-evolution of the Characteristic Frequency}
\label{subsec:frequency_coevolution}

\begin{figure}[t]
    \centering
    \includegraphics[width=\columnwidth]{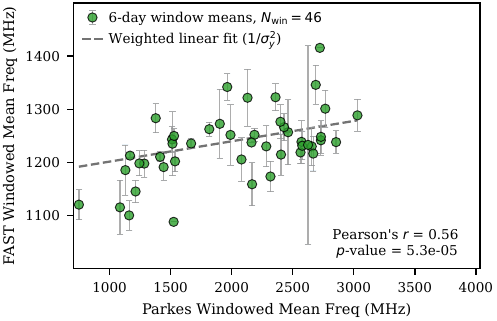}
    \caption{
    Comparison of the 6-day binned mean frequencies measured by Parkes/Murriyang and FAST during overlapping epochs.
    Each point represents a non-overlapping 6-day MJD bin with valid measurements from both telescopes.
    The axis ranges indicate the observing bands of Parkes/Murriyang (704--4032 MHz) and FAST (1000--1500 MHz).
    The grey dashed line shows the weighted linear fit using FAST frequency uncertainties.
    }
    \label{fig:fast_parkes_freq}
\end{figure}

Using Parkes/Murriyang UWL observations, \citet{2026arXiv260512098L} reported a periodic modulation of the burst central frequency with a period of approximately $112.91$~days and a positive Parkes--FAST frequency correlation.
However, their FAST comparison sample extended only to MJD~60552, while FAST monitoring continued to MJD~60832.
Our burst-cluster catalog provides average peak-frequency measurements for drifting FAST burst clusters with $N_{\rm sub}\geq2$, which show substantial variations throughout the campaign and extend the FAST frequency coverage for comparison with Parkes.
We divided the jointly covered FAST--Parkes interval into non-overlapping 6-day MJD bins and calculated the mean frequency in each bin.
After retaining bins with valid measurements from both telescopes, 45 paired bins remain (Figure~\ref{fig:fast_parkes_freq}).
The two datasets show a significant positive Pearson correlation ($r=0.56$, $p=5.3\times10^{-5}$).
Despite differences in instrumental bandpass and frequency estimation, FAST and Parkes show broadly consistent long-term evolution of the characteristic emission frequency.

\subsection{Burst-cluster duration and waiting-time distributions}
\label{sec:duration_waiting_time}

Figure~\ref{fig:overview}(c) shows the burst-cluster duration for $N_{\rm sub}\geq2$ clusters.
The daily median and geometric mean show no clear secular trend.
The time-integrated distribution is strongly asymmetric, with a median of $15.7$~ms, an interquartile range of $4.5$-$67.1$~ms, and linear- and log-space means of $57.04$ and $17.51$~ms, respectively, reflecting the extended long-duration tail.

\begin{figure}[t!]
    \centering
    \includegraphics[
        width=\columnwidth,
        height=\textheight,
        keepaspectratio
    ]{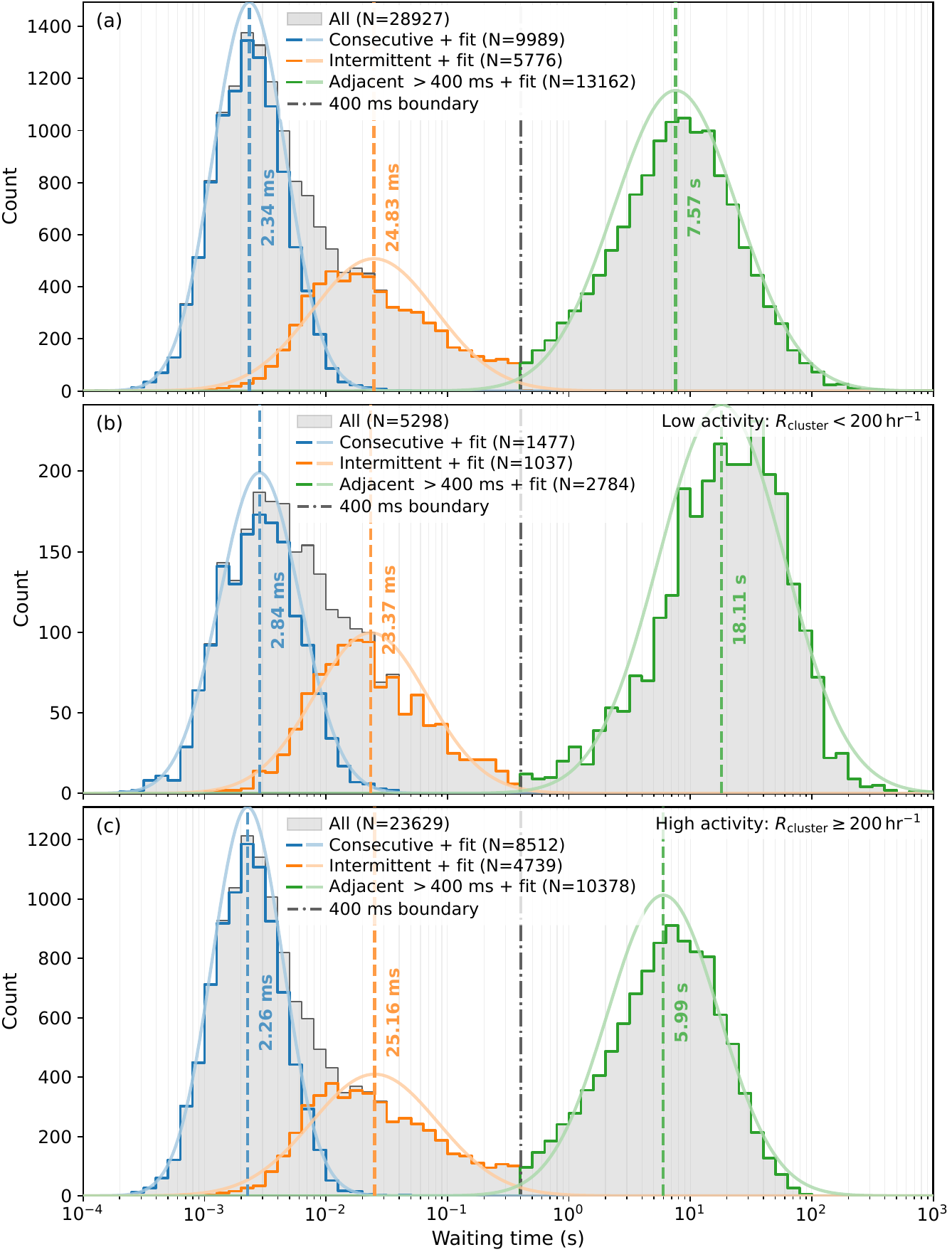}
    \caption{
    Waiting-time distributions of FRB~20240114A.
    Panel (a) shows the full sample, while panels (b) and (c) show the low- and high-activity samples, respectively, separated at a burst-cluster rate of $200~{\rm hr^{-1}}$.
    The grey histogram shows the total waiting-time distribution, while the colored histograms show the intra-cluster (consecutive and intermittent) and inter-cluster ($>400$~ms) components.
    Solid curves show the log-normal fits, colored dashed lines mark the characteristic timescales $e^\mu$, and the grey dash-dotted line indicates the $400$~ms selection threshold.
    }
    \label{fig:waiting_time_distribution}
\end{figure}

Figure~\ref{fig:waiting_time_distribution}(a) combines three types of waiting time: consecutive time intervals calculated between adjacent sub-bursts within each burst, intermittent intervals calculated between adjacent bursts within each burst cluster, and inter-cluster intervals obtained from the topocentric MJD differences between adjacent BurstIDs within each continuous observing session after retaining only intervals longer than $400$~ms.
The three samples contain 9989, 5776, and 13162 intervals, with medians of $2.31$~ms, $21.73$~ms, and $7.92$~s and interquartile ranges of $1.43$-$3.69$~ms, $10.22$-$55.57$~ms, and $3.46$-$16.90$~s, respectively.
The characteristic timescale of each log-normal component is defined as $e^\mu$.
The resulting timescales are $2.34$~ms, $24.83$~ms, and $7.57$~s, corresponding to separations between sub-bursts, bursts, and burst clusters.
Similar distinctions between intra-burst and inter-burst waiting times have been adopted in previous studies \citep{2023MNRAS.519..666J,2026A&A...709A..47L}.
A multi-component waiting-time structure is also found in a separate collaborative study by \citet{2026arXiv260819713Z}, based on the earlier FAST burst catalog of \citet{2025arXiv250714707Z}.
Panels~\ref{fig:waiting_time_distribution}(b) and (c) show the corresponding distributions for low- and high-activity sessions, respectively, where we empirically adopt a burst-cluster rate of \(200~{\rm hr^{-1}}\) as an operational criterion to separate different observed activity levels.
The fitted mean values for the first and second components are 2.84 ms vs. 2.26 ms, and 23.37 ms vs. 25.16 ms, which are very close to each other, while the third component is 18.11 s vs. 5.99 s, which is very different.
This indicates the shorter two components are likely intrinsic, while the longer one is more or less related to the event rate.
Notice the dividing line of high activity and low activity is set as $200 ~{\rm hr}^{-1} = 1/18 ~{\rm s}^{-1}$, which also roughly divides these two mean values of waiting time of inter-clusters between high active and low active ones. 

To test the inter-cluster arrival process, we followed the Weibull formalism of \citet{2018MNRAS.475.5109O} and the session-resolved procedure of \citet{2026A&A...709A..47L}.
The waiting-time density is
\begin{equation}
W(\delta\mid k,r)
=
\frac{k}{\delta}
A^{k}\mathrm{e}^{-A^{k}},
\label{eq:weibull_waiting_time}
\end{equation}
where $A=\delta r\Gamma\left(1+1/k\right)$, $\delta$ is the inter-cluster waiting time, $r$ is the mean burst-cluster rate within a given observing session, and $k$ is the Weibull shape parameter.
The case $k=1$ corresponds to a Poisson process, whereas $k<1$ and $k>1$ indicate more temporally clustered and more uniformly spaced burst-cluster arrivals, respectively, than expected for a Poisson process.

Because only intervals with $\delta>\delta_0=0.4$~s were retained, we used the left-truncated density
\begin{equation}
W_{\rm tr}(\delta\,\vert\,k,r,\delta_0)
=
\frac{W(\delta\,\vert\,k,r)}
{S(\delta_0\,\vert\,k,r)},
\qquad
\delta>\delta_0,
\label{eq:truncated_weibull_waiting_time}
\end{equation}
where $S(\delta_0\,\vert\,k,r)$ is the Weibull survival probability at the truncation boundary.
This likelihood explicitly accounts for the $400$~ms selection threshold.

\begin{figure*}[t!]
    \centering
    \includegraphics[
        width=0.90\textwidth,
        height=0.68\textheight,
        keepaspectratio
    ]{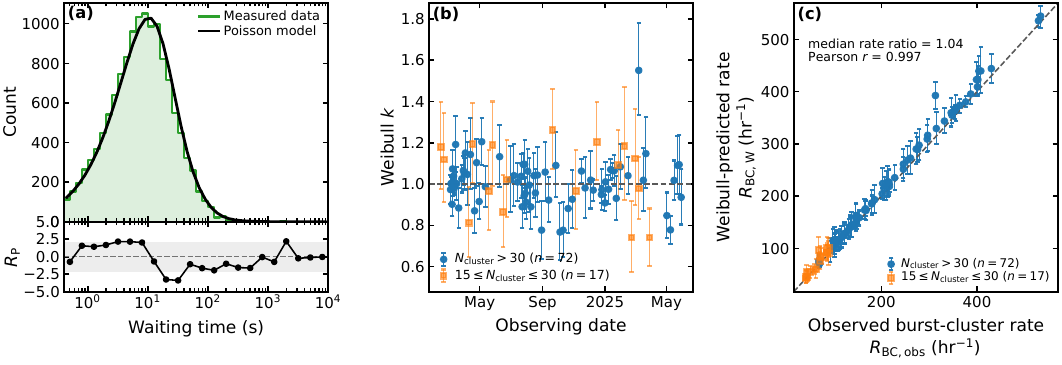}
    \caption{
    Session-resolved Weibull and Poisson analysis of inter-cluster intervals longer than $0.4$~s.
    Panel (a) compares 13162 observed intervals with the stacked conditional Poisson model, with the lower strip showing Pearson residuals and the grey region marking $\left|R_{\rm P}\right|\leq2$.
    Panel (b) shows the posterior median and $16$th-$84$th percentile interval of $k$ relative to the Poisson value $k=1$ (dashed line).
    Blue circles and orange squares denote sessions with $N_{\rm cluster}>30$ and $15\leq N_{\rm cluster}\leq30$, respectively.
    Panel (c) compares Weibull-predicted and measured burst-cluster rates, with the dashed line denoting equality.
    }
    \label{fig:weibull_poisson_summary}
\end{figure*}

Figure~\ref{fig:weibull_poisson_summary}(a) compares the 13162 long intervals with a stacked session-dependent Poisson model.
Setting $k=1$ in Equation~(\ref{eq:weibull_waiting_time}) and applying the same left truncation at $\delta_0$ gives the exponential density $r\mathrm{e}^{-r(\delta-\delta_0)}$ for $\delta>\delta_0$.
For each observing session $j$, we adopted the directly measured burst-cluster rate, $R_{\mathrm{BC},j}=N_{\mathrm{cluster},j}/T_{\mathrm{obs},j}$, and combined the corresponding distributions as
\begin{equation}
W_{\mathrm{P,stack}}(\delta)
=
\sum_j
w_j
R_{\mathrm{BC},j}
\mathrm{e}^{-R_{\mathrm{BC},j}(\delta-\delta_0)},
\label{eq:stacked_poisson_waiting_time}
\end{equation}
where $w_j=N_{\mathrm{int},j}/N_{\mathrm{int}}$, $N_{\mathrm{int}}=\sum_j N_{\mathrm{int},j}$, and $\delta>\delta_0$.
The curve in panel (a) is therefore the weighted sum of the truncated Poisson waiting-time distributions calculated from the measured burst-cluster rate of each observing session.
The agreement between the observed and predicted counts is quantified by the Pearson residual, $R_{\rm P}=(O-E)/\sqrt{E}$, where $O$ and $E$ are the observed and predicted counts in each bin.

For the session-resolved Weibull analysis shown in Figure~\ref{fig:weibull_poisson_summary}(b), each continuous observing session with at least three retained inter-cluster intervals (97/99 sessions) was fitted independently, treating each unique Burst-ClusterID as one event and fitting $k$ and $r$ jointly.
The reported $k$ values are the medians of the marginalized posteriors, with 16th--84th percentiles as $1\sigma$ credible intervals.
The joint $(r,k)$ and marginalized posterior distributions will be released with the data products described in the Data Availability section.
All retained intervals connect different Burst-ClusterIDs, with no pairs separated by $\leq400$~ms.
Following the criterion of \citet{2026A&A...709A..47L}, we define the primary sample as 72 sessions with $N_{\rm cluster}>30$ and the secondary sample as 17 sessions with $15\leq N_{\rm cluster}\leq30$.
Among the primary sample, 50, 69, and 71 sessions are consistent with $k=1$ within $1\sigma$, $2\sigma$, and $3\sigma$, respectively.
Only the 2025 May 9 session deviates beyond $3\sigma$, with $k=0.778^{+0.069}_{-0.069}$.
Given the number of tested sessions, this isolated deviation should be interpreted cautiously and does not by itself provide strong evidence for persistent non-Poissonian clustering.
Panel (c) compares Weibull-predicted and measured burst-cluster rates.
Together with the predominance of $k\approx1$, these results indicate that the long inter-cluster intervals are broadly consistent with a rate-variable Poisson process.

\subsection{Drifting-rate stability and its dependence on spectral properties}
\label{sec:drifting_rate}

\begin{figure*}[t]
    \centering
    \includegraphics[width=\textwidth]{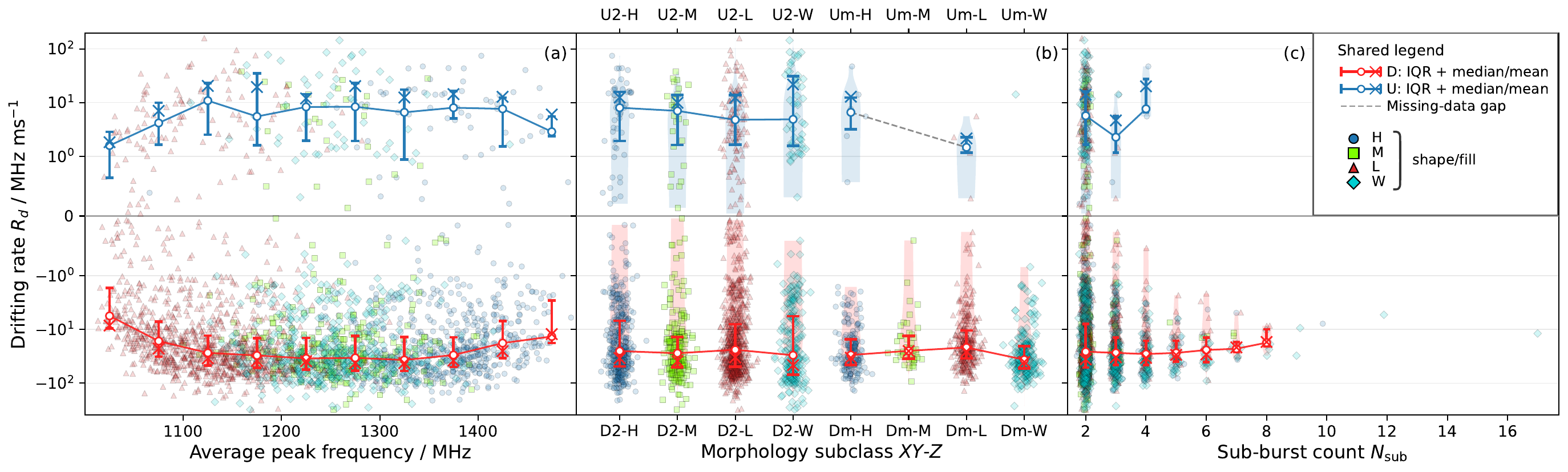}
    \caption{
    Drifting-rate relations for FRB~20240114A.
    Panels (a)--(c) show $R_{\rm d}$ as a function of average peak frequency, morphological subclass, and sub-burst count $N_{\rm sub}$, respectively.
    Symbols and fill colors distinguish the H/M/L/W subclasses.
    Summary markers show the interquartile range, median, and mean; grey dashed lines indicate missing-data gaps.
    }
    \label{fig:drift_rate}
\end{figure*}

Figure~\ref{fig:overview}(e) shows the long-term evolution of the drifting rate for 249 upward-drifting U2/Um burst clusters and 2318 downward-drifting D2/Dm burst clusters.
Despite burst-to-burst scatter, daily geometric means show no secular trend over the monitoring campaign.
The time-integrated upward-drifting distribution has a median of $5.6~\mathrm{MHz\,ms^{-1}}$ and a 25th--75th percentile range of $1.6$--$15.7~\mathrm{MHz\,ms^{-1}}$, whereas the downward-drifting distribution has a median of $-26.2~\mathrm{MHz\,ms^{-1}}$ and a corresponding range of $-48.7$ to $-10.0~\mathrm{MHz\,ms^{-1}}$.
Figure~\ref{fig:drift_rate} compares the intra-cluster drifting rate $R_{\rm d}$, defined as the frequency--time drift among sub-bursts within each cluster, with average peak frequency, morphology, and sub-burst count.
Panel (a) shows that the magnitude of $R_{\rm d}$ tends to decrease as the average peak frequency approaches the band edges.
This may result from partial spectral coverage, which underestimates the drifting-rate magnitude.
Measurements near the band center ($\nu_{\rm pk}\simeq1250~\mathrm{MHz}$) are less affected.
Panel (b) compares drifting-rate distributions among morphological subclasses.
For D2-M ($N=215$) and U2-M ($N=26$), the median drifting rates are $-27.7$ and $7.0~\mathrm{MHz\,ms^{-1}}$, with interquartile ranges of $-50.4$ to $-13.9~\mathrm{MHz\,ms^{-1}}$ and $1.6$ to $13.9~\mathrm{MHz\,ms^{-1}}$, respectively.
Panel (c) shows only weak dependence of drifting rates on $N_{\rm sub}$.

\subsection{Candidate Simultaneous Sub-bursts}
\label{sec:simultaneous_subbursts}

\begin{figure*}[t]
    \centering
    \includegraphics[width=\textwidth]
    {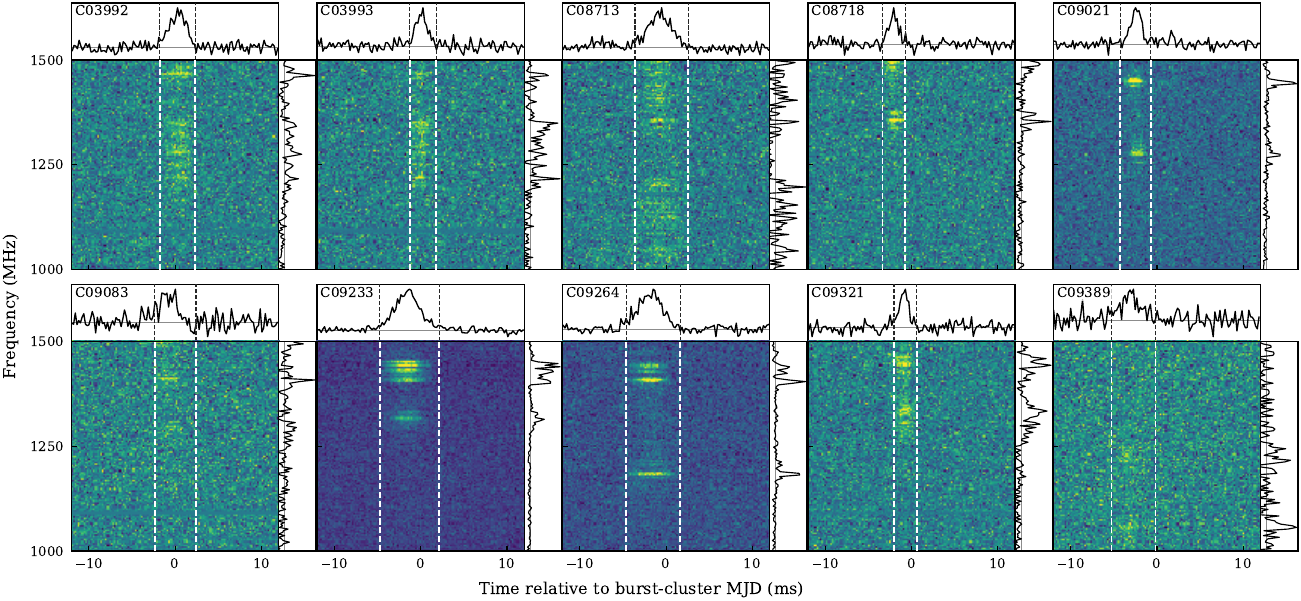}
    \caption{
    Representative events showing two or more frequency-separated features that appear nearly simultaneous in the dynamic spectrum.
    The upper profile is the frequency-integrated light curve, the lower image is the dedispersed and downsampled dynamic spectrum, and the profile on the right is integrated over the effective width marked by the dashed lines.
    }
    \label{fig:simultaneous_subbursts}
\end{figure*}

Several events contain two or more frequency-separated emission regions that appear nearly simultaneous (Figure~\ref{fig:simultaneous_subbursts}).
It remains unclear whether these regions represent distinct sub-bursts or structured emission within a single sub-burst.
A frequency gap consistent with the noise level does not necessarily indicate separate emission components, as weak connecting emission may remain below the detection threshold.
Moreover, scintillation or other frequency-dependent propagation effects may produce apparently separated spectral regions from intrinsically continuous emission.
To ensure consistent measurements of consecutive time intervals between sub-bursts, these nearly simultaneous frequency-separated regions are treated as a single sub-burst.
Their physical origin remains uncertain and is being investigated in a separate collaborative theoretical study by J.-H.~Cao et al.\ (in preparation).

\section{Burst Energetics}
\label{sec:energetics}

\subsection{Energy Dependence on Burst-Cluster Multiplicity and Activity State}
\label{sec:energy_rate}

\begin{figure*}[t]
    \centering
    \includegraphics[width=\textwidth]{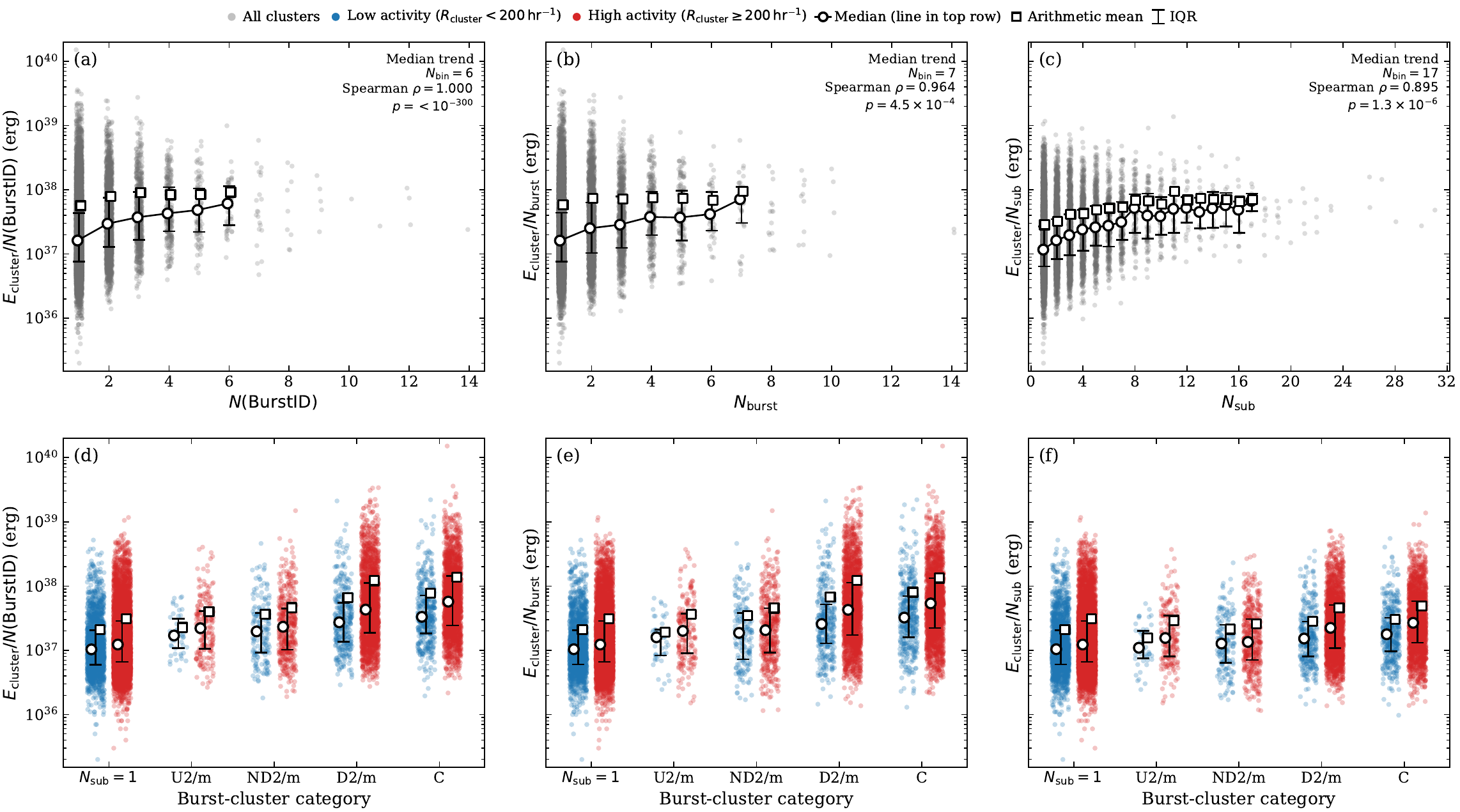}
    \caption{
    Energy dependence on burst-cluster multiplicity and activity state.
    Panels (a)--(c) show $E_{\mathrm{cluster}}/N(\mathrm{BurstID})$, $E_{\mathrm{cluster}}/N_{\mathrm{burst}}$, and $E_{\mathrm{cluster}}/N_{\mathrm{sub}}$ as functions of counts.
    Grey points represent burst clusters, while open circles, open squares, and error bars denote the median, arithmetic mean, and 25th--75th percentile range, respectively.
    The median values are connected, and Spearman correlations are calculated from count-bin medians over $N(\mathrm{BurstID})\leq6$, $N_{\mathrm{burst}}\leq7$, and $N_{\mathrm{sub}}\leq17$.
    Panels (d)--(f) compare energy-per-count measures among clusters with $N_{\mathrm{sub}}=1$, U2/m, ND2/m, D2/m, and C classifications.
    Blue and red points correspond to sessions with $R_{\mathrm{cluster}}<200~\mathrm{hr}^{-1}$ and $R_{\mathrm{cluster}}\geq200~\mathrm{hr}^{-1}$, respectively.
    }
    \label{fig:energy_count}
\end{figure*}

We characterize the internal multiplicity of each burst cluster using three count-based measures: the number of catalogue BurstIDs, $N(\mathrm{BurstID})$, the number of temporally separated bursts, $N_{\mathrm{burst}}$, and the total number of resolved sub-bursts, $N_{\mathrm{sub}}$.
For each cluster, the total energy, $E_{\mathrm{cluster}}$, is obtained by summing the energies of all matched BurstIDs within that cluster.
We calculate energy per BurstID, burst, and sub-burst using $E_{\mathrm{cluster}}/N(\mathrm{BurstID})$, $E_{\mathrm{cluster}}/N_{\mathrm{burst}}$, and $E_{\mathrm{cluster}}/N_{\mathrm{sub}}$, respectively.
In the full burst-cluster sample, the median energy per counted unit increases monotonically or nearly monotonically with all three multiplicity measures, as shown in the upper panels of Figure~\ref{fig:energy_count}.
Over $N(\mathrm{BurstID})\leq6$, $N_{\mathrm{burst}}\leq7$, and $N_{\mathrm{sub}}\leq17$, the Spearman coefficients calculated from the count-bin medians are $\rho=1.000$, $0.964$, and $0.895$, respectively.
Despite broad scatter, the increasing medians indicate that characteristic energy per counted unit increases with burst-cluster multiplicity.

The lower panels of Figure~\ref{fig:energy_count} compare the three energy-per-count measures among clusters with $N_{\mathrm{sub}}=1$, U2/m, ND2/m, D2/m, and C classifications.
For all morphology categories and energy definitions, high-activity sessions show higher median and arithmetic-mean energies per counted unit than low-activity sessions.
Across both activity states, C and D2/m clusters have the highest characteristic energies, with C ranking first and D2/m second, followed by U2/m and ND2/m.
Clusters with $N_{\mathrm{sub}}=1$ show the lowest energies per BurstID and burst in panels (d) and (e), while their energy per sub-burst is comparable to U2/m and ND2/m in panel (f).

\subsection{Energy--DM Relation across Burst-Cluster Morphologies}
\label{sec:energy_dm_relation}
\begin{figure*}[t]
    \centering
    \includegraphics[width=\textwidth]{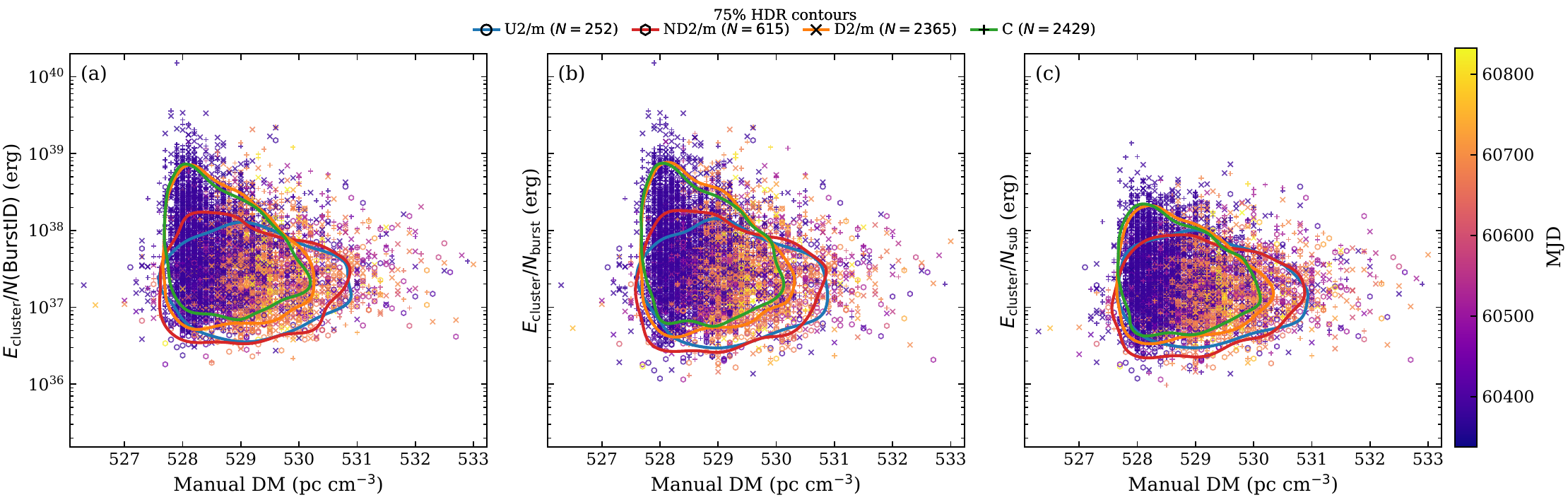}
    \caption{
    Cluster-averaged energy measures as functions of Manual DM for different morphology groups.
    Panels (a)--(c) show $E_{\mathrm{cluster}}/N(\mathrm{BurstID})$, $E_{\mathrm{cluster}}/N_{\mathrm{burst}}$, and $E_{\mathrm{cluster}}/N_{\mathrm{sub}}$, respectively.
    Each point represents a burst cluster, with symbols denoting morphology groups and colors indicating topocentric MJD.
    The sample contains 5661 clusters (U2/m, ND2/m, D2/m, and C: 252, 615, 2365, and 2429 clusters).
    Contours enclose the 75\% highest-density regions (HDRs) of each morphology group.
    }
    \label{fig:energy_dm}
\end{figure*}

To explore whether the morphology-dependent energetic differences are accompanied by variations in other measured physical quantities, we examined both the manually refined DM and the independently measured RM reported by \citet{2026ApJS..284...77W} across the morphology groups.
No clear morphology-dependent trend was found in RM, whereas a more apparent pattern is seen in the joint distribution of energy and Manual DM.
We select burst clusters in the U2/m, ND2/m, D2/m, and C categories, all of which have Manual DM measurements.
The resulting sample contains 5661 clusters in the U2/m, ND2/m, D2/m, and C categories, with 252, 615, 2365, and 2429 clusters, respectively.
Figure~\ref{fig:energy_dm} shows $E_{\mathrm{cluster}}/N(\mathrm{BurstID})$, $E_{\mathrm{cluster}}/N_{\mathrm{burst}}$, and $E_{\mathrm{cluster}}/N_{\mathrm{sub}}$ as functions of Manual DM.
The 75\% highest-density regions (HDRs) of the four morphology groups overlap substantially, but a broad pairwise tendency is visible across all three energy definitions.
C and D2/m occupy the higher-energy part of the distributions and tend toward slightly lower and more concentrated Manual DM values, whereas U2/m and ND2/m occupy lower characteristic-energy levels and extend toward somewhat higher and broader Manual DM ranges.

\section{Discussion and Conclusion}
\label{sec:Discussion}

The long-term FAST monitoring reveals substantial activity variations in FRB~20240114A, with the sub-burst, burst, and burst-cluster rates evolving broadly in parallel, indicating that the observed activity variations are robust against different levels of temporal grouping.
The Manual DM shows a gradual secular increase at a rate of \(1.29\pm0.11~\mathrm{pc\,cm^{-3}\,yr^{-1}}\), consistent with the independently reported value of \(1.10\pm0.07~\mathrm{pc\,cm^{-3}\,yr^{-1}}\) \citep{2026arXiv260518513O}.
At the population level, \citet{2026ApJ..1007..173C} and \citet{2026arXiv260508410C} reported statistically significant secular DM increases for two repeating FRBs, with typical rates of DM increase of \(\sim1.1~\mathrm{pc\,cm^{-3}\,yr^{-1}}\), comparable to our result.

As shown in Figure~\ref{fig:energy_count}, the burst energetics of FRB~20240114A vary systematically with both activity state and burst-cluster morphology.
The higher characteristic energies observed during high-activity sessions are qualitatively consistent with the positive log-linear relation between burst rate and mean burst energy reported by \citet{2026arXiv260702939N}, despite differences in the energy definitions and temporal selections between the two studies.
The energy distributions also show a partial grouping by morphology, with C and D2/m clusters occupying higher characteristic-energy levels than U2/m and ND2/m. 
The tendency for higher-multiplicity clusters to exhibit larger characteristic energies suggests that burst complexity may be related to the energy release process, although the substantial scatter prevents a simple one-to-one correspondence.
Recent statistical studies of multicomponent burst structures in repeating FRBs have revealed characteristic distributions of component numbers, indicating that burst morphology itself contains statistical information about the emission process \citep{2026ApJ...996....4Z}.
Figure~\ref{fig:energy_dm} shows a corresponding broad pairwise tendency in the Manual DM--energy plane.
However, the substantial overlap among the morphology groups, together with the secular evolution of Manual DM, means that this possible morphology--energy--DM connection should be interpreted with caution.

FRB~20240114A exhibits different frequency-evolution behavior across temporal scales. Within individual burst clusters, downward drifting strongly dominates over upward drifting, qualitatively consistent with the preference for negative frequency changes at millisecond-to-sub-second separations reported by \citet{2026A&A...709A..47L}. The measured drifting-rate magnitudes tend to decrease toward the edges of the FAST observing band, where incomplete spectral coverage may lead to underestimation, while their dependence on \(N_{\rm sub}\) is weak. The observed drift characteristics may also be affected by propagation effects such as plasma lensing \citep{LiRN2026}.

At longer waiting times, extending from sub-second scales to a few hundred seconds, positive and negative values of \(B\) become approximately symmetric and the arrival times are consistent with Poisson statistics, indicating neither a preferred direction of frequency change nor an apparent tendency toward temporal clustering or regular spacing among more widely separated events \citep[Figure~7]{2026A&A...709A..47L}.
Our session-resolved Weibull analysis independently shows that the inter-cluster intervals are broadly compatible with a rate-variable Poisson process, as shown in Figure~\ref{fig:weibull_poisson_summary}.
Similar scale-dependent behavior has been reported for other active repeaters. FRB~20121102A shows strongly clustered arrival times when closely separated events are included, but becomes consistent with a Poisson process after excluding a small number of intervals shorter than one second \citep{2018MNRAS.475.5109O,2021MNRAS.500..448C}, while FRB~20220912A exhibits distinct short- and long-timescale waiting-time components, with some observing sessions having Weibull shape parameters consistent with $k=1$ \citep{2024MNRAS.534.3331K}.
The combination of an approximately Poissonian arrival process and the absence of a preferred inter-cluster frequency direction is also consistent with treating the \(0.4\)~s valley in our waiting-time distribution as the boundary between intra-cluster structure and separate burst clusters.
These results suggest that each burst cluster may represent a distinct emission episode and that successive episodes are approximately independent within a continuous observing session.
Such episodes could be repeatedly produced by the same physical emission region or originate from different regions within the source.
The present observations do not distinguish between these possibilities, and weak correlations between successive burst clusters cannot be excluded.

On timescales of days to months, the Parkes/Murriyang monitoring campaign provides a long baseline for tracing the characteristic emission frequency \citep{2026arXiv260216409U}.
Using this sample, \citet{2026arXiv260512098L} identified a \(112.91\)-day frequency modulation at a significance exceeding \(6\sigma\), found no corresponding periodicity in the burst arrival times, and reported a positive correlation with the FAST measurements available at the time of that study, which extended only to MJD~60552.
Our burst-cluster catalog extends the FAST frequency coverage to MJD~60832, and the FAST and Parkes/Murriyang characteristic frequencies remain significantly correlated, with \(r=0.56\) and \(p=5.3\times10^{-5}\), as shown in Figure~\ref{fig:fast_parkes_freq}.

The burst-cluster catalog presented here provides a uniform framework for investigating this multi-timescale behavior, while the retained \texttt{BurstID} indexing enables direct integration with complementary burst-level measurements. Our principal conclusions are summarized as follows:
\begin{enumerate}[
    topsep=0.4\baselineskip,
    itemsep=0.3\baselineskip,
    parsep=0pt,
    partopsep=0pt
]
\item The 495-day active episode contains 13261 burst clusters, 19037 bursts, and 29026 sub-bursts detected in 99 observing sessions.
The Manual DM shows a gradual secular increase of \(1.29\pm0.11~\mathrm{pc\,cm^{-3}\,yr^{-1}}\), enlarging the still limited but growing sample of repeaters with increasing DM.

\item FRB~20240114A exhibits hierarchical spectral--temporal behavior across different timescales.
At the shortest timescales, sub-bursts within burst clusters show a strong preference for downward drifting over upward drifting, with a number ratio of approximately \(9.4\!:\!1\).
At intermediate timescales, separate burst clusters show inter-cluster arrival times broadly consistent with Poisson statistics, suggesting that they may represent approximately independent emission episodes within observing sessions, while previously reported frequency changes between separate burst clusters show no clear preferred direction.
On longer timescales, the characteristic emission frequency exhibits evolution consistent with the long-term modulation revealed by Parkes/Murriyang.

\item Burst energetics show systematic trends with both burst-cluster multiplicity and activity state.
The characteristic energy generally increases with burst-cluster multiplicity and is systematically higher during high-activity sessions across morphological categories.

\item The DM--energy distributions show a weak morphology-dependent tendency.
The higher-energy C and D2/m groups tend to have slightly lower and more concentrated Manual DMs, but the substantial overlap among groups suggests that this association remains tentative.

\end{enumerate}

Overall, the FAST burst-cluster catalog of FRB~20240114A provides a comprehensive view of repeating FRB emission across multiple timescales and physical properties.
These results highlight the complex and hierarchical nature of repeating FRB activity, while a unified physical framework remains to be established.

\section*{Data Availability}
\hypertarget{sec:DataAvailability}{}

The dataset underlying this work is publicly available via the Science Data Bank (\url{https://doi.org/10.57760/sciencedb.fastro.00013}) and provides the burst-cluster catalog of FRB~20240114A.
\begin{acknowledgments}
This work made use of the data from FAST FRB Key Science Project. This work is supported by National Natural Science Foundation of China (NSFC) Programs Nos. 12603071, 12588202, 12041306, 11988101, 11725313, 11690024, 12041303, 12421003, U1731238, 12233002, 12522305, 12303042, 12403100, W2442001, 12503055, 12203045, 12447115, U1931203, 12533001, 12473001, 12503050, and 12575049; CAS International Partnership Program No. 114-A11KYSB20160008; CAS Strategic Priority Research Program No. XDB23000000; the National SKA Program of China Nos. 2022SKA0130100.
W.W.Z. is supported by CAS Project for Young Scientists in Basic Research, YSBR-063.
D.L. is a New Cornerstone investigator.
P.W. acknowledges support from the CAS Youth Interdisciplinary Team, the Youth Innovation Promotion Association CAS (id. 2021055), and the Cultivation Project for FAST Scientific Payoff and Research Achievement of CAMS-CAS.
R.L. is supported by the Guangdong Provincial Natural Science Foundation (Grant No. 2026A1515030042).
Y.F. is supported by the Leading Innovation and Entrepreneurship Team of Zhejiang Province of China grant No. 2023R01008, and by Key R\&D Program of Zhejiang grant No. 2024SSYS0012.
Y.F.H. acknowledges the support from the Xinjiang Tianchi Program and the National Key R\&D Program of China (2021YFA0718500).
Q.W. is supported by the China Postdoctoral Science Foundation (CPSF) under Grant Nos. GZB20240308, 2025T180875.
C.W.T. is supported by CAS project No. JZHKYPT-2021-06.
J.R.N. is supported by the Postdoctoral Fellowship Program of CPSF under Grant No. GZB20250737.
J.J.C. acknowledges financial support from the European Research Council (ERC) starting grant ``COMPACT" (Grant agreement number 101078094).
Y.H.Q. is supported by the Research Council of Finland Centre of Excellence in Neutron-Star Physics (project 374063).
X.Z. acknowledges support from the National SKA Program of China (Grant Nos. 2022SKA0110200 and 2022SKA0110203).
This work made use of data from FAST, a Chinese national mega-science facility built and operated by the National Astronomical Observatories, Chinese Academy of Sciences.
This research has made use of the CHIME/FRB VOEvent Service, BlinkVerse and TransientVerse.
The computation was partially completed on the HPC Platform of Huazhong University of Science and Technology.
We thank Marlon Bause for helpful comments.
We thank Meng-Pei Wang and Rui Min, both astronomy enthusiasts, for their assistance with a small part of the data processing.

\end{acknowledgments}

\facilities{FAST}

\software{
TransientX \citep{2024A&A...683A.183M},
PSRCHIVE \citep{2004PASA...21..302H}
}

\bibliography{references}{}
\bibliographystyle{aasjournalv7}



\end{document}